\documentclass[conference]{IEEEtran}

\IEEEoverridecommandlockouts
\usepackage{cite}
\usepackage{amsmath,amssymb,amsfonts}
\usepackage{algorithmic}
\usepackage{graphicx}
\usepackage{textcomp}
\usepackage{xcolor}
\def\BibTeX{{\rm B\kern-.05em{\sc i\kern-.025em b}\kern-.08em
    T\kern-.1667em\lower.7ex\hbox{E}\kern-.125emX}}
    
\usepackage[utf8]{inputenc}
\usepackage{amsmath,amssymb,amsfonts}
\usepackage{algorithmic}
\usepackage{bm}
\usepackage{algorithm}
\usepackage{graphicx}
\usepackage{textcomp}
\usepackage{xcolor}
\usepackage{listings}
\usepackage{xargs}
\usepackage{xcolor}
\usepackage{todonotes}
\usepackage{tabularx}
\usepackage{booktabs}
\usepackage{float}
\usepackage{enumitem}
\usepackage{setspace}
\usepackage{multicol}
\usepackage{multirow}
\usepackage{mwe}
\usepackage{siunitx}
\usepackage{subcaption}
\newfloat{algorithm}{t}{lop}

\newcolumntype{Y}{>{\centering\arraybackslash}X}

\newcommandx{\todoitem}[2][1=]{\todo[linecolor=blue,backgroundcolor=blue!25,bordercolor=blue,#1]{#2}}

\newcommand{\discriminatorperf}{59.3}
\newcommand{\generatorperf}{38.6}

\newcommand{\heropeakperf}{1228}

\newcommand{\herosustainedperf}{1207}

\newcommand{\weakscalingeff}{93.1}

\newcommand{\heropeakperfpar}{1134}
\newcommand{\herosustainedperfpar}{1107}
\newcommand{\weakscalingeffpar}{91.6}

\newcommand{\heropeakperfover}{1228}
\newcommand{\herosustainedperfover}{1136}
\newcommand{\weakscalingeffover}{93.7}

\newcommand{\heropeakperfseq}{604}
\newcommand{\herosustainedperfseq}{597}
\newcommand{\weakscalingeffseq}{92.0}

\newcommand{\heronodecount}{4584}
\newcommand{\herogpucount}{27,500}

\begin{document}


\title{Highly-scalable, physics-informed GANs for learning solutions of stochastic PDEs}

\author{
 
 \IEEEauthorblockN{Liu Yang\IEEEauthorrefmark{1},
 Sean Treichler\IEEEauthorrefmark{2},
 Thorsten Kurth\IEEEauthorrefmark{3},
 Keno Fischer\IEEEauthorrefmark{4},\\
 David Barajas-Solano\IEEEauthorrefmark{5},
 Josh Romero\IEEEauthorrefmark{2},
 Valentin Churavy\IEEEauthorrefmark{6},
 Alexandre Tartakovsky\IEEEauthorrefmark{5},\\
 Michael Houston\IEEEauthorrefmark{2},
 Prabhat\IEEEauthorrefmark{3},
 George Karniadakis\IEEEauthorrefmark{1}
 }
 
 \IEEEauthorblockA{
 \IEEEauthorrefmark{1}
 Brown University, Providence, RI 02912, USA
 \{liu\_yang,george\_karniadakis\}@brown.edu}
 
 \IEEEauthorblockA{
 \IEEEauthorrefmark{2}
 NVIDIA,
 Santa Clara, CA 95051, USA
 \{sean,joshr,mhouston\}@nvidia.com}
 
 \IEEEauthorblockA{
 \IEEEauthorrefmark{3}
Lawrence Berkeley National Laboratory, Berkeley, CA 94720, USA
 \{tkurth,prabhat\}@lbl.gov}
 
 \IEEEauthorblockA{
 \IEEEauthorrefmark{4}
 Julia Computing, Cambridge, MA 02139, USA
 keno@juliacomputing.com}
 
 \IEEEauthorblockA{
 \IEEEauthorrefmark{5}
 Pacific Northwest National Lab, Richland, WA 99352, USA
 \{david.barajas-solano,alexandre.tartakovsky\}@pnnl.gov}
 
 \IEEEauthorblockA{
 \IEEEauthorrefmark{6}
 Massachusetts Institute of Technology, Cambridge, MA 02139, USA
 vchuravy@mit.edu}
 
 }

\maketitle
\begin{abstract}



  Uncertainty quantification for forward and inverse problems is a central challenge across physical and biomedical disciplines. 
  We address this challenge for the problem of modeling subsurface flow at the Hanford Site by combining stochastic computational models with observational data using physics-informed GAN models. 
  The geographic extent, spatial heterogeneity, and multiple correlation length scales of the Hanford Site require training a computationally intensive GAN model to thousands of dimensions. 
  We develop a hierarchical scheme
  for exploiting domain parallelism, map discriminators and generators to multiple GPUs, and employ efficient communication schemes to ensure training stability and convergence.
  We developed a highly optimized implementation of this scheme that scales to {\herogpucount} NVIDIA Volta GPUs and {\heronodecount} nodes on the Summit supercomputer with a \weakscalingeff\% scaling efficiency, achieving peak and sustained half-precision rates of {\heropeakperf} PF/s and {\herosustainedperf} PF/s. 

\end{abstract}




\section{Overview}
\label{sec:science-drivers}


\subsection{Parameter estimation and uncertainty quantification for subsurface flow models}


Mathematical models of subsurface flow and transport are inherently uncertain because of the lack of data about the distribution of geological units, the distribution of hydrological properties (e.g., hydraulic conductivity) within each unit, and initial and boundary conditions.
Here, we focus on parameterization and uncertainty quantification (UQ) in the subsurface flow model at the Department of Energy's Hanford Site, one of the most contaminated sites in the western hemisphere. During the Hanford Site's 60-plus years history,
there have been more than 1000 individual sources of contaminants
distributed over 200 square miles mostly along Columbia River~\cite{thorne2006groundwater}. Accurate subsurface flow models with rigorous UQ are necessary for assessing risks of the contaminants reaching the Columbia river and numerous wells used by agriculture and as sources of drinking water, as well as for the design of efficient remediation strategies.  

\subsection{UQ with Stochastic Partial Differential Equations}

Uncertain initial and boundary conditions and model parameters render the governing model equations stochastic. In this context, UQ becomes equivalent to solving stochastic PDEs (SPDEs). Forward solution of SPDEs requires that 
all model parameters as well as the initial/boundary conditions are prescribed either deterministically or stochastically, which is
not possible unless experimental data are available to provide additional information for critical parameters, e.g. the field conductivity. The Hanford Site has about 1,000 to 10,000 wells equipped with sensors that can provide 
diverse sources of additional information. To this end, we can formulate -- instead of a forward problem -- a {\it mixed} problem, wherein we fuse observational data and mathematical models into one unified formulation. The solution of the mixed problem will then yield both the quantity of interest (e.g., pressure head) as well as the conductivity fields in the form of stochastic processes. 

\subsection{UQ with physics-informed GANs (PI-GANs)}

Generative adversarial networks (GANs) can learn probability distributions from data, and hence they can be employed for modeling the {\em stochasticity} in physical systems. In order to encode the physics into GANs,  we follow the physics-informed neural network approach~\cite{MaziarParisGK17-1,MaziarParisGK17-2}, which was subsequently implemented using GANs for 1D problems in~\cite{YangLiu-GAN}.
The authors of~\cite{YangLiu-GAN} demonstrated that WGANs-GP, in particular, can represent stochastic processes accurately and can approximate solutions of SPDEs in high dimensions. Specifically, it was found that when doubling the number of stochastic dimensions (from 30 to 60), the computational cost is approximately doubled,
suggesting that the overall computational cost increases with a low-polynomial growth.

\begin{figure}[h]
  \includegraphics[width=\linewidth]{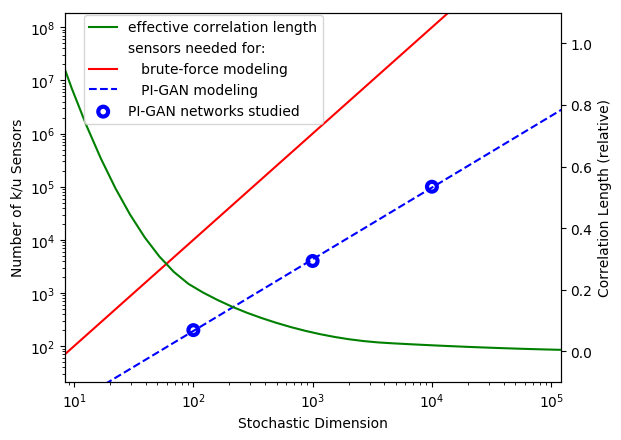}
  \caption{Visualization of relationships between stochastic dimension, correlation length, and the rough sensor counts required for brute-force modeling of k/u fields vs. learning distributions using a PI-GAN.   The three networks considered in this paper are indicated with blue circles.}
  \label{fig:cartoon}
\end{figure}

This observation is illustrated in Figure~\ref{fig:cartoon}.  The green line shows how stochastic dimension increases
rapidly as the correlation distance of the behavior of interest shrinks.  (Some modeling problems related to
the Hanford Site have relative correlation distances on the order of one part per million.)  A brute-force attempt to
model the stochastic processes in a 2-D system without knowledge of the physics would require (at least) a
quadratically-increasing number of sensors (the red line in the figure), whereas PI-GANs can, in principle,
feasibly tackle very high dimensional stochastic problems.  The blue line in Figure~\ref{fig:cartoon} shows their
expected growth rate, and in this work, we study three specific PI-GAN network configurations indicated by the
blue circles.

\subsection{Contributions}
Tackling a realistic dataset from the Hanford Site (corresponding to a large stochastic dimensionality and number of sensors) requires development of a scalable PI-GAN framework that can obtain high levels of performance on computational nodes, and scale across massive HPC systems. Towards accomplishing this goal, our paper makes the following contributions:
\begin{itemize}

\item{Development and validation of physically-informed GAN architecture for modeling subsurface flow. First demonstration of framework to problem with unprecedented stochastic dimensionality (1,000). }

\item{A novel domain decomposition algorithm for GANs
    involving one generator and hundreds of discriminators.}

\item{Highly optimized generator and discriminator architecture implementation in TensorFlow of the proposed domain decomposition algorithm that obtains {\discriminatorperf} TF/s and {\generatorperf} TF/s on Volta GPUs}

\item{Highly scalable architecture 
that obtains \weakscalingeff\% scaling efficiency up to {\heronodecount} Summit nodes.}

\item{Demonstrated \textbf{{\herosustainedperf} PF/s sustained} and \textbf{{\heropeakperf} PF/s peak half-precision (FP16) performance} for distributed training of PI-GAN architecture on {\herogpucount} Volta GPUs.}

\end{itemize}

\section{State of the Art}
\label{sec:intro}


\subsection{State-of-the-art in SPDE solvers}

The full solution of a system of SPDEs is the joint probability density function (PDF) of the state variables. The most common method for UQ is  Monte Carlo simulation (MCS)~\cite{wasserstein1997monte}, a straightforward and robust but computationally demanding sampling technique. Generalized Polynomial Chaos (PC)~\cite{wiener1938homogeneous,XiuKarniadakis-2003} methods have emerged in the past two decades as an effective method for solving SPDEs in low to moderate dimensions. All PC methods share the same foundational principles: (i) approximating infinite-dimensional random parameter fields with finite-term expansions (where the number of terms in the finite-term expansion defines the dimensionality of the probability space); and (ii) adopting stochastic Galerkin  method or  choosing collocation points in the probability space to compute moments of the PDFs~\cite{elman2011assessment}. As a result, they all suffer from the so-called ``curse of dimensionality'':
their computational cost increases exponentially with the number of stochastic dimensions. The state-of-the-art PC methods, including enhancements like ANOVA, become less computationally efficient than MCS for approximately 100 random dimensions~\cite{JasmineFoo,Handy-SISC,barajassolano-2016-stochastic}. Based on available data, we estimate that the heterogeneity and multiscale dynamics of the Hanford Site requires at least 1,000 random dimensions for adequate representation, which is out of reach of the PC methods. In addition, the PC methods are most efficient for computing the leading moments of the joint PDFs, i.e., the mean describing the average behavior of the system, and the variance measuring the scale of uncertainty. While mean and variance are sufficient for describing uncertainty in some systems, many applications require full PDFs, especially those involving risk analysis or those monitoring rare but possibly catastrophic events. We hypothesize that PI-GANs have the potential to address both challenges; they can mitigate the curse of dimensionality and provide an accurate estimate of the entire PDF of the parameters and states rather than leading moments.

Here it must be noted that in comparison to PC methods, MCS does not suffer from the curse of dimensionality and thus can be employed for UQ in highly complex stochastic problems if large-scale computational resources are available.
On the other hand, MCS cannot be used to assimilate measurements (which is one of the main objectives of this work) and must be coupled with other methods such Ensemble Kalman filter~\cite{evensen2003ensemble}. 
The comparative advantages and disadvantages of physics-informed learning and PDE-based methods for parameter estimation are discussed in the next section. 


\subsection{State-of-the-art in physics-informed Deep Learning}

The specific data-driven approach to modeling multiscale physical systems depends crucially on the amount of data available as well as on the complexity of the system itself. For modeling the Hanford Site, we assume that we know the physics partially, but we have several scattered measurements  that we can use to infer the missing functional terms and other parameters in the PDE and simultaneously recover the solution. This is the {\it mixed} case that we address in this paper, but in a significantly more complex scenario, that is, where the solution is a stochastic process due to stochastic extrinsic excitation or an uncertain material property, e.g. permeability or diffusivity in a porous medium. Physics-informed learning can be realized by two different approaches: Using either Gaussian Process Regression (GPR) that employs informed priors based on the PDE that expresses the physical law~\cite{Maziar-GPR}, or based on deep neural networks (DNNs) that encode the PDE, which induces a very complex network sharing hyperparameters with the primary uniformed network that deals with labeled data~\cite{Maziar-PINNs}. 

Including the PDE directly into the DNN has several advantages, since the constrained PDE, which is part of the loss function, acts as a regularization term. For example, in \cite{tartakovsky2018learning} it was demonstrated that physics-informed neural networks (PINNs) learn the accurate conductivity field $k$, while the Tikhonov regularization term of the standard maximum a posteriori (MAP) parameter estimation method causes nonphysical perturbations in $k$.
Moreover, the optimization problem is relatively easier to solve as it is constrained on the solution manifold. Compared to classical numerical solvers, there is no need of a structured mesh since all derivatives are computed using automatic differentiation while the PDE residual penalized in the loss function is evaluated at random points in the space-time domain. No classical discretizations are employed, so we avoid the numerical diffusion and dispersion errors that are harmful for propagation phenomena. However, we still have to deal with the optimization, generalization, and approximation errors as in all standard DNNs.

Most published work on physics-informed neural networks deals with deterministic systems, but there have been a few papers published on data-driven methods for SPDEs, e.g., \cite{Weinan-arxiv, Maziar-arxiv, Paris-encoder}, for forward problems. For the mixed and inverse problems, Zhang et al~\cite{zhang2018quantifying} have recently proposed a DNN based method that learns the modal functions of the quantity of interest, which could also be the unknown system parameters. The effectiveness of PINNs is demonstrated in Fig. 11 of~\cite{zhang2018quantifying} where the k-sensors at four different locations
have all the same value but, unlike a data-driven-only DNN, the PINN-based method is able to reproduce the correct variation of k everywhere in the domain.
While robust, this method does suffer from the ``curse of dimensionality'' in that the number of PC expansion terms grows exponentially as the effective dimension increases, leading to prohibitive computational costs for modeling stochastic systems in high dimensions.
Here, we follow the preliminary work of ~\cite{YangLiu-GAN} to design a physics-informed GAN in order to approximate solutions of an elliptic stochastic PDE as we present below. 

\subsection{State-of-the-art in GANs}

Generative Adversarial Networks (GANs) are generative models originally proposed by Goodfellow et al.~\cite{goodfellow2014generative}.  The basis of a GAN is two neural networks playing an adversarial game.  The generator  tries to produce data from a probability distribution function in an attempt to trick the discriminator.  The discriminator tries to assess if the data that it has been presented with is "fake" (i.e from the generator) or "real" (i.e from the training dataset). Over the course of the training, the discriminator teaches the generator how to produce realistic output.  By the end of a successful training process, the generator should learn to generate the proper data distribution and the discriminator would be completely fooled in assessing whether the generator's output is real or fake.


GANs are an active area of research with several variants adapted for different problems. Some of the more popular work is in the image synthesis such as DCGAN~\cite{dcgan-paper-2015}, and StyleGAN~\cite{stylegan}. 
In Wasserstein GANs with clips on weights (weight-clipped WGANs)~\cite{arjovsky2017wasserstein} and WGANs with gradient penalty (WGAN-GP)~\cite{gulrajani2017improved}, the loss for the generator has a meaningful mathematical interpretation, namely the Wasserstein distance between generated distribution and real distribution. In \cite{YangLiu-GAN}, the authors illustrated that WGAN-GP is more suitable for learning stochastic processes compared with vanilla GANs, especially for cases with deterministic boundaries. 

One property shared by all current GAN variants is brittleness when attempting to scale out
training to a large number of computational nodes.  Recent efforts have succeeded in pushing the feasible data-parallel batch
size for GANs to 2048\cite{largegan}, but that is still an order of magnitude below what would be needed to scale
to the Summit system and its 27,600 GPUs.  Our effort relies on the use of several forms of parallelism, limiting
the reliance on data-parallelism and keeping global batch sizes well below 1000.

\section{Innovations}
\label{sec:innovations}


\subsection{Problem Setup}

We consider two-dimensional depth-averaged steady-state saturated flow at the Hanford Site described by the combination of the Darcy Law and continuity equation,
\begin{equation}
  \label{Darcy}
  \nabla \cdot [k(x) \nabla u(x)]=f(x),
\end{equation}
where $k(x)$ is the transmissivity or depth-averaged hydraulic conductivity, $u(x)$ is the hydraulic head, and $f(x)$ is the source term (i.e., infiltration from the vadose zone).
Close to ten thousand measurements of $k$ and/or $u$ are available from pumping well experiments, direct observations of water level in the wells, and other geophysical measurements, including electrical resistivity data.
The estimated $k(x)$ and the steady-state boundary conditions for the Hanford Site were reported in~\cite{thorne2006groundwater,cole2001uncertainty}.
The same publications found significant uncertainty in the measured conductivity values and the estimated parameters.
To quantify uncertainty, we treat $k(x)$ in Eq. (\ref{Darcy}) as a random field and, following the common practice in geostatistics, we assume that $k(x)$ has a lognormal distribution,
\begin{equation}
  \label{mean}
  k(x) = \exp \left [ \overline{Y}(x) + Y'(x) \right ] =K_g(x) \exp \left [Y'(x) \right ],
\end{equation}
with a synthetic mean $\overline{Y}(x)$, and $Y'(x)$ given by a zero-mean Gaussian process with prescribed covariance kernel.
For this study we employ the isotropic, stationary, covariance kernel
\begin{equation}
  \label{covariance}
  \overline{Y'(x) Y'(y)} = \sigma^2
  \exp \left ( |x_1 - y_1| / \lambda + |x_2 - y_2| / \lambda \right ).
\end{equation}
where $\sigma$ and $\lambda$ denote the standard deviation of $\log(k)$ fluctuations and their correlation length, respectively.
 
The $u$ measurements at the Hanford Site are constantly collected in a number of wells.
The boundary conditions (river stages) have seasonal variations that we disregard in this study, and instead focus on a single season when the boundary conditions fluctuate around their mean values.
This leads to the variations in of $u$ measurements recorded in the wells.
The objective of this work is to learn the joint distributions of $k(x)$ and $u(x)$ that (i) follow the $k(x)$ model in Eqs.~(\ref{mean}) and (\ref{covariance}); (ii) match $u$ measurements; and (iii) satisfy the physical law in Eq.~(\ref{Darcy}). 

\subsubsection{Datasets}


\begin{table}[ht]
  \centering
  \small
  \begin{tabularx}{\columnwidth}{|l|cccYY|}
  \hline
    Dataset & $\lambda$ & Stoc Dim & Levels & No. of subdomains & No. of $f$, $\log(k)$--$u$ sensors \\ \hline
    1 & \num{0.5} & \num{100} & \num{1} & \num{1} & \num{50}, \num{50} \\
    2 & \num{0.5} & \num{100} & \num{2} & $[\num{1}, \num{99}]$ & \num{100}, \num{100} \\
    3 & \num{0.5} & \num{100} & \num{2} & $[\num{1}, \num{19}]$ & \num{100}, \num{100} \\
    4 & \num{0.3} & \num{1000} & \num{2} & $[\num{1}, \num{19}]$ & \num{100}, \num{100} \\
    5 & \num{0.3} & \num{1000} & \num{2} & $[\num{1}, \num{19}]$ & \num{2000}, \num{100} \\
    6 & 0.1467 & \num{10000} & \num{3} & $[\num{1}, \num{19}, \num{480}]$ & \num{2000}, \num{100} \\ \hline
  \end{tabularx}
  \caption{Datasets used in this study.}
  \label{tab:datasets}
\end{table}

To learn the joint density of $k(x)$ and $u(x)$, we generate ensembles of $\num{1e5}$ synthetic realizations of $k(x)$, and for each realization we solve Eq.~(\ref{Darcy}) for $u(x)$ with $f(x) = 0$.
The $u(x)$ solutions are computed by discretizing the Hanford Site geometry into a regular grid of 503,937 cells and employing a cell-centered finite volume scheme together with the two-point flux approximation.

The $k(x)$ realizations are generated by sampling the GP model~(\ref{mean}) and (\ref{covariance}).
For all our experiments we set $\sigma = 0.5$.
To learn the capacity of the proposed approach to learn different correlation structures of $\log(k)$, we generate various ensembles with dimensionless correlation lengths $\lambda = \num{0.5}$, $\num{0.3}$, and $\num{0.1467}$ (that is, relative to the Hanford Site lengthscale of $60$ \si{km}.
The resulting datasets are listed in Table~\ref{tab:datasets}.
For each correlation length, the fluctuations $Y'(x)$ are sampled using the corresponding Karhunen-Lo\'{e}ve (KL) expansion truncated to retain $98\%$ to $99\%$ of the variance $\sigma^2$.
For decreasing correlation length, the number of KL terms, or {\it stochastic dimension}, increases.
The resulting number of stochastic dimensions are listed in Table~\ref{tab:datasets}.
We note that for Dataset \# 6, the number of realizations is $\num{1e6}$.

\begin{figure}[h]
  \includegraphics[width=0.7\linewidth]{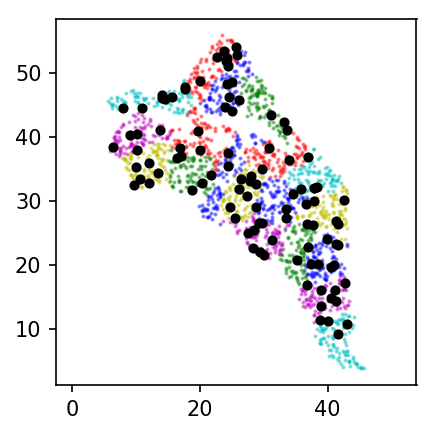}
  \caption{Locations of $\log(k)$ and $u$ sensors for levels 1 (black) and 2 (color).
  Units are in \si{\km}.}
  \label{fig:hanford-1500-k-u-sensors-lv-0-1}
\end{figure}

To train the PI-GAN model at scale, we partition the training datasets into a hierarchy of spatial subdomains, as described in Section~\ref{sec:baseline-architecture}.
At each level of the hierarchy, the FV cells of the discretization are partitioned into well-balanced subdomains using METIS~\cite{karypis1998metis}.
For each subdomain, the locations of $\log(k)$ and $u$ sensors are chosen randomly from the subdomain FV cells; these locations are fixed for each dataset.
On the other hand, $f$ sensor locations are chosen randomly for each realization in each dataset.
The number of levels, subdomains per level, $\log(k)$ and $u$ sensors, and $f$ sensors for each dataset are listed in Table~\ref{tab:datasets}.
Figure~\ref{fig:hanford-1500-k-u-sensors-lv-0-1} shows the $\log(k)$ and $u$ sensor locations for levels 1 and 2 of Datasets \#3 to \#6.

\subsection{Physics-Informed GANs}


\subsubsection{Baseline Architecture}
\label{sec:baseline-architecture}

\begin{figure}[h]
  \includegraphics[width=\linewidth]{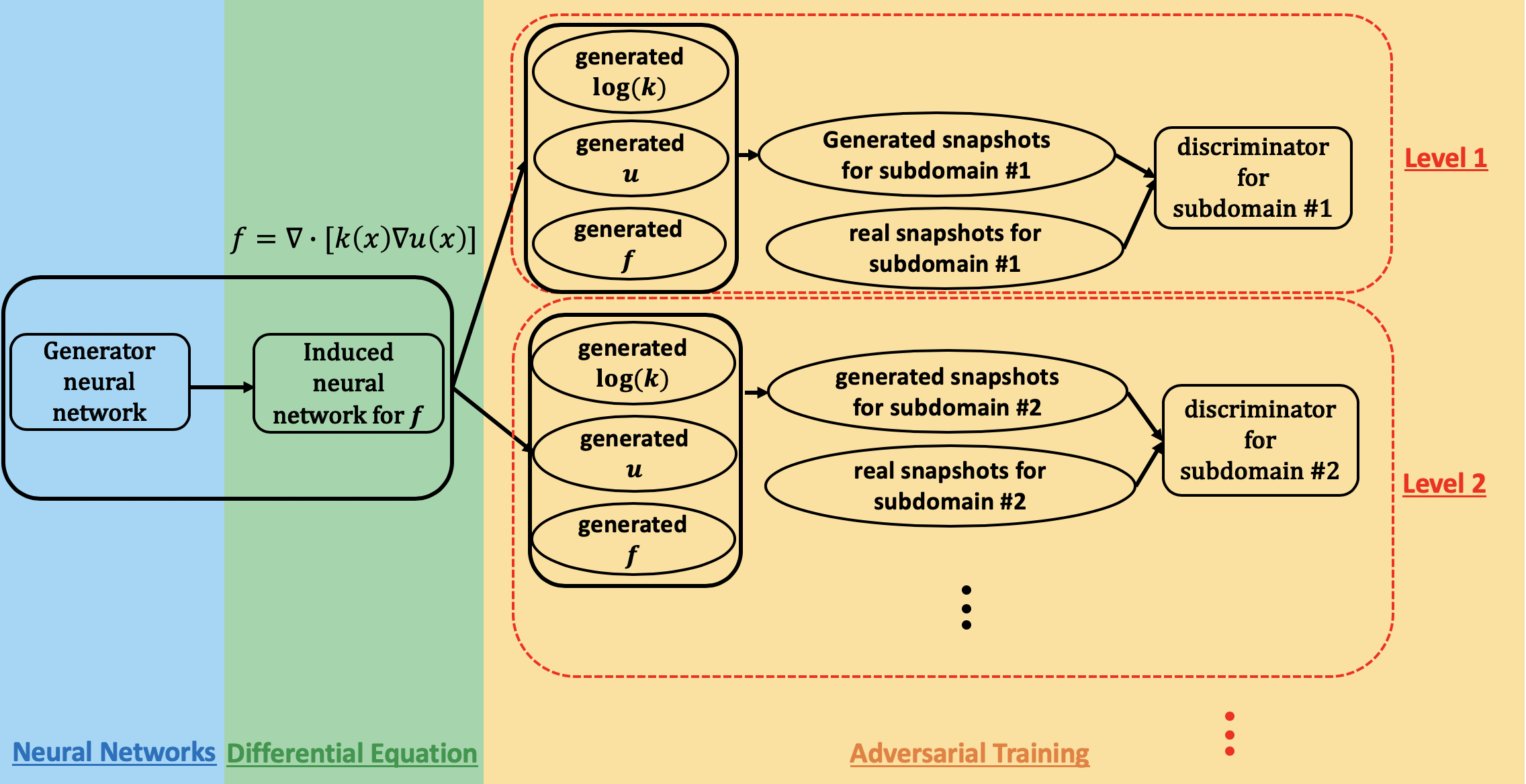}
  \caption{PI-GAN architecture used in the study.}
  \label{fig:pi-gan-schematic}
\end{figure}
Figure \ref{fig:pi-gan-schematic} depicts the PI-GAN architecture used in this study.
The generator neural network takes a random vector and spatial coordinates as inputs to represent $\log(k)$ and $u$. Using automatic differentiation, we can apply the differential operator to the generator neural network and get an induced neural network to represent $f$ analytically. To learn the joint distribution of $\log(k)$ and $u$ with the constraint of the PDE~(Eq. (\ref{Darcy})), for a fixed input random vector, we generate the $\log(k)$ and $u$ values at positions of all sensors, as well as $f$ values at random positions scattered in the domain (we call them free sensors), and concatenate these values and the corresponding spatial coordinates as one ``fake'' snapshot. We feed the ensemble of ``fake'' snapshots for various input random vectors and the ensemble of ``real'' snapshot into another deep neural network, namely the discriminator, and train the discriminator and generator in an adversarial fashion.

Note that the input size to the discriminator is linear in the total number of sensors of $\log(k)$, $u$ and $f$. As the number of sensors becomes large, it could become difficult to train the discriminator. To tackle this we introduce the notion of \emph{domain decomposition} for PI-GANs. While we still have one generator for all the subdomains, we partition the whole domain into subdomains and assign one discriminator for each subdomain. Subsequently, each discriminator will only be in charge of all or part of the sensors within the assigned subdomain. Moreover, in order to learn the long range and short range correlation simultaneously, we arrange the subdomains in a hierarchical structure. For example, we can have one subdomain at level 1, 19 subdomains at level 2, and 480 subdomains at level 3 \emph{all partitioning the whole domain}.
Figure~\ref{fig:hanford-1500-k-u-sensors-lv-0-1} illustrates a sample level 1 and level 2 partitioning scheme.
As the level decreases, the area of each subdomain increases while the sensors become more sparse, so that the total number of sensors in each subdomain will be limited and balanced. The lower level subdomains and discriminators are in charge of the relatively long range correlation, while the higher level subdomains and discriminators are in charge of relatively short range correlation, providing higher resolution of the field. For each subdomain, we feed the generated and real snapshots at the corresponding sensors to the discriminator assigned to this subdomain. The loss of the generator is the weighted average of negative critics from all the discriminators. To balance the levels, in practice for certain subdomain at a  certain level we set the weight as 1 divided by the product of the number of levels and the number of subdomains in this level.

\subsubsection{Diagnostics}
The objective of this work is to learn the distribution of the parameter $k$ and $u$ at sensors, and predict the whole fields of $k$ and $u$ with the information from PDE~(Eq. (\ref{Darcy})). Since the distribution of $\log(k)$ is close to normal, we focus on validation of $\log(k)$ instead of $k$ directly.
The predictions of $\log(k)$ and $u$ are given in terms of their mean (the most probable value) and standard deviation (the measure of uncertainty). Moreover, to validate the distributions of the learned random field $\log(k)$, we compare its (learned) correlation with those of the $\log(k)$ GP model, assumed here to be the ground truth. We also compare the matching between the pairwise covariance and $L_1$ distance between validation points in $\log(k)$ field, in order to give a more detailed illustration of the correlation of learned $\log(k)$ field. We will revisit these diagnostics for evaluating PI-GANs in Section \ref{sec:science_results}.



\subsection{Single Node optimizations}
\label{sec:innov:singlenode}

We implement and train our PI-GAN networks in TensorFlow~\cite{tensorflow2015-whitepaper, tensorflow-site}, a framework with a CUDA backend capable of utilizing GPUs for the intense computations required during the training process.
The core of both the generator and discriminator networks is a series of fully-connected layers in which both
the number of layers and the width of the layers is parameterized.  We focus on three configurations in our
effort: $5{\times}128$ (i.e. 5 layers, each with 128 neurons), $11{\times}512$, and $21{\times}1024$.  (These
correspond to the blue circles in Figure~\ref{fig:cartoon}.)  As is common for deeper neural
networks, the design includes {\em residual} connections that skip intermediate layers to improve gradient
propagation and trainability~\cite{resnet}.  The generator network must generate both $\log(k(x))$ and $u(x)$, and we
chose a design that uses a single set of weights (thus reducing the memory footprint) and produces
separate $\log(k(x))$ and $u(x)$ outputs in the final layer, ensuring that TensorFlow's automatic (numeric) differentiation transformation
does not introduce unnecessary computation (i.e., we need $\nabla u(x)$ but not $\nabla \log(k(x))$).

An important difference
between the two networks exists in how the inputs are provided.  The generator network is designed with a narrow
input that operates on a single sensor location (along with the random noise input), forcing each $\log(k(x))$ to be
generated independently.  The many sensors from a single realization are provided as separate samples in an
effectively-larger batch, mitigating the inefficiency of the small batch sizes used for the generator (due
to memory constraints).  In contrast, the discriminator network is given all the sensor locations for its subdomain
as a single input, allowing it to directly learn correlations between different sensor locations.  This is
intended to allow the discriminator to train more efficiently, which in turn drives the generator network to do
a better job.  Additionally, the differences between the generator and discriminator setups discourage
"common-mode" failures that can occur with GANs where the discriminator simply mirrors the generator rather
than learning properties of the target distribution.

TensorFlow includes support for mixed-precision training.  We employ the standard methodology of computing
activations and gradients in half precision (FP16) but accumulating gradients and updating model weights in
single precision (FP32)~\cite{nvidia:mixedprectraining}.  The use of adaptive loss scaling\cite{nvidia:mixedprectraining} maximizes the
effectiveness of the relatively small dynamic range of the half precision format.

A final design choice lies in the selection of the non-linear activation functions that are inserted between the
fully-connected layers in the network.  The most common (and best optimized in frameworks and GPU libraries)
function is ReLU\cite{relu}, but its merely piecewise-differentiability is a concern for a network that 
incorporates differentiation in the forward pass as well as backpropagation.  We instead chose the
{\em hyperbolic tangent} (tanh) activation function for its differentiability and unbiasedness, despite an expected
performance cost (see Section~\ref{sec:perfsingle}).

\subsection{Multi-node optimizations}

\subsubsection{Parallelism}
\label{sec:parallelism}
The standard technique used for scaling out a deep learning workload is {\em data-parallelism}, where GPUs
work with identical copies of the entire network and each computes gradients for a separate set
(the {\em local batch}) of input samples.  A summation is performed across all the GPUs to produce the gradients
for the {\em global batch}, and each GPU then makes identical updates to their copies of the weights, ensuring the
model stays consistent across the entire system.  Software libraries such as Horovod~\cite{sergeev2018horovod}
simplify the incorporation of data-parallelism with a few lines of additional code.  Unfortunately, a purely data-parallel
approach for our PI-GAN runs into three challenges at scale:
\begin{enumerate}
    \item At least part of the gradient summation performed in each training step represents exposed communication (i.e. it cannot be overlapped with any computation).  As the number of GPUs involved in the summation grows,
    the latency of this summation unavoidably increases, reducing the overall throughput.
    \item The quality of a trained model can suffer if the global batch size becomes too large.  Although some classification networks have been trained with batch sizes up to 64K\cite{Tencent}, GANs have proven to be more
    sensitive, and the best known techniques have only been able to push to a global batch size of 2048\cite{largegan}.  Even with a local
    batch size of 1, a purely data-parallel run on all of Summit would use a global batch size over 27,000.
    \item Finally, the model for our PI-GAN simply doesn't fit within a single GPUs device memory.  Again, even
    with a local batch size of 1, our medium network (designed for a stochastic dimension of 1000) would require
    over 19 GB of memory (once all intermediate results for back-propagation are accounted for) while our larger network would require over 1.7 TB of memory (on each GPU).
\end{enumerate}
We address all of these challenges by using a novel hybrid parallelism technique for our PI-GAN.  We first divide
the pool of available GPUs into two subgroups.  The {\em generator workers} are responsible for training the
generator model against the collection of discriminator models, while each {\em discriminator worker} contributes
to the training of one of the discriminators.  Our design permits an adjustable ratio of generator workers to discriminator workers, which is one of the hyperparameters that impacts the relative training rates.  (Somewhat conveniently, a 50/50 split works well for our largest network). Both subgroups are then further divided by
associating each worker with a single subdomain.  If care is taken to balance the work per subdomain (as shown
in Figure~\ref{fig:hanford-1500-k-u-sensors-lv-0-1}), these divisions are uniform, but uneven splits can be made
if the subdomain sizes are imbalanced.
The discriminator workers assigned to a given subdomain then use traditional data-parallelism to train the
subdomain's discriminator.  At full Summit scale, this corresponds to a global batch size of ${\sim}350$, well 
within the feasibility of data-parallel GAN training. In contrast, the generator workers assigned to a given
subdomain do not operate independently.  Instead, they use a form of {\em model-parallelism}, computing only
the portion of the generator loss function associated with
their assigned subdomain.  The losses from each subdomain must be summed together, but a rearrangement of terms
based on the linearity of the gradient computation allows this summation to be folded into the data-parallel
summation for stochastic gradient descent (up to a constant multiplicative factor, which is easily compensated
for):
\begin{equation*}
  \nabla \ell \approx \frac{1}{N} \sum_i^N \nabla \ell(x_i) =
  \frac{1}{N} \sum_{(i,d)}^{N\times D} \nabla \ell_d(x_i)
\end{equation*}
Horovod supports the ability to restrict its gradient reduction to an MPI subcommunicator.  By carefully splitting
the job into $D+1$ sub-communicators (1 for the generator training and 1 for each discriminator subdomain), the
training of each model continues to benefit from the simplicity and performance of the Horovod approach.

\subsubsection{Model Exchange}
\label{sec:schedules}

The parallelism technique described above splits the training of different models across different GPUs, but the
adversarial approach of a GAN requires that the generator train against recent discriminator models and vice-versa.
These model exchanges must occur outside the subcommunicators created for Horovod, and are performed using the
{\tt mpi4py} Python bindings for MPI.  We explored and implemented several different {\em schedules} for model exchange, which we
describe next and evaluate in Section~\ref{sec:scaling-results}.

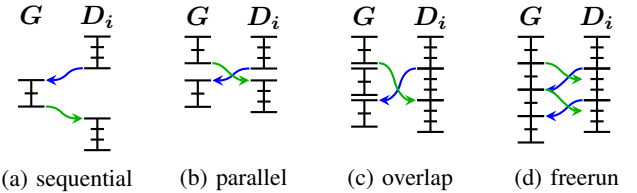
\begin{figure}[ht]
  \begin{subfigure}{0.25\linewidth}
    \centering
    \begin{tikzpicture}[x=0.5em,y=-0.5em]
      \draw[white] (2.5,0) -- (2.5,8.6);
      \node[anchor=north] at (0,-3) {$\bm{G}$};
      \begin{scope}[shift={(0,3.3)}]
        \draw[thick] (-1,0) -- (1,0);
        \draw[thick] (-0.5,1) -- (0.5,1);
        \draw[thick] (-1,2) -- (1,2);
        \draw[thick] (0,0) -- (0,2);
      \end{scope}
      \node[anchor=north] at (5,-3) {$\bm{D_i}$};
      \begin{scope}[shift={(5,0)}]
        \draw[thick] (-1,0) -- (1,0);
        \draw[thick] (-0.5,0.8) -- (0.5,0.8);
        \draw[thick] (-0.5,1.6) -- (0.5,1.6);
        \draw[thick] (-1,2.4) -- (1,2.4);
        \draw[thick] (0,0) -- (0,2.4);
      \end{scope}
      \begin{scope}[shift={(5,6.2)}]
        \draw[thick] (-1,0) -- (1,0);
        \draw[thick] (-0.5,0.8) -- (0.5,0.8);
        \draw[thick] (-0.5,1.6) -- (0.5,1.6);
        \draw[thick] (-1,2.4) -- (1,2.4);
        \draw[thick] (0,0) -- (0,2.4);
      \end{scope}
      \draw[thick,blue,->,>=stealth] (3.9,2.4) to [out=180,in=45] (2.7,2.85) to [out=225,in=0] (1.1,3.3);
      \draw[thick,black!30!green,->,>=stealth] (1.1,5.3) to [out=0,in=135] (2.3,5.75) to [out=315,in=180] (3.9,6.2);
    \end{tikzpicture}
    \caption{sequential}
    \label{fig:sched:seq}
  \end{subfigure}%
  \begin{subfigure}{0.25\linewidth}
    \centering
    \begin{tikzpicture}[x=0.5em,y=-0.5em]
      \draw[white] (2.5,0) -- (2.5,8.6);
      \node[anchor=north] at (0,-3) {$\bm{G}$};
      \begin{scope}[shift={(0,0)}]
        \draw[thick] (-1,0) -- (1,0);
        \draw[thick] (-0.5,1) -- (0.5,1);
        \draw[thick] (-1,2) -- (1,2);
        \draw[thick] (0,0) -- (0,2);
      \end{scope}
      \begin{scope}[shift={(0,3.3)}]
        \draw[thick] (-1,0) -- (1,0);
        \draw[thick] (-0.5,1) -- (0.5,1);
        \draw[thick] (-1,2) -- (1,2);
        \draw[thick] (0,0) -- (0,2);
      \end{scope}
      \node[anchor=north] at (5,-3) {$\bm{D_i}$};
      \begin{scope}[shift={(5,0)}]
        \draw[thick] (-1,0) -- (1,0);
        \draw[thick] (-0.5,0.8) -- (0.5,0.8);
        \draw[thick] (-0.5,1.6) -- (0.5,1.6);
        \draw[thick] (-1,2.4) -- (1,2.4);
        \draw[thick] (0,0) -- (0,2.4);
      \end{scope}
      \begin{scope}[shift={(5,3.3)}]
        \draw[thick] (-1,0) -- (1,0);
        \draw[thick] (-0.5,0.8) -- (0.5,0.8);
        \draw[thick] (-0.5,1.6) -- (0.5,1.6);
        \draw[thick] (-1,2.4) -- (1,2.4);
        \draw[thick] (0,0) -- (0,2.4);
      \end{scope}
      \draw[thick,blue,->,>=stealth] (3.9,2.4) to [out=180,in=45] (2.7,2.85) to [out=225,in=0] (1.1,3.3);
      \draw[thick,black!30!green,->,>=stealth] (1.1,2.0) to [out=0,in=135] (2.3,2.65) to [out=315,in=180] (3.9,3.3);
    \end{tikzpicture}
    \caption{parallel}
    \label{fig:sched:par}
  \end{subfigure}%
  \begin{subfigure}{0.25\linewidth}
    \centering
    \begin{tikzpicture}[x=0.5em,y=-0.5em]
      \draw[white] (2.5,0) -- (2.5,8.6);
      \node[anchor=north] at (0,-3) {$\bm{G}$};
      \begin{scope}[shift={(0,0)}]
        \draw[thick] (-1,0) -- (1,0);
        \draw[thick] (-0.5,1) -- (0.5,1);
        \draw[thick] (-1,2) -- (1,2);
        \draw[thick] (0,0) -- (0,2);
      \end{scope}
      \begin{scope}[shift={(0,2.4)}]
        \draw[thick] (-1,0) -- (1,0);
        \draw[thick] (-0.5,1) -- (0.5,1);
        \draw[thick] (-1,2) -- (1,2);
        \draw[thick] (0,0) -- (0,2);
      \end{scope}
      \begin{scope}[shift={(0,4.8)}]
        \draw[thick] (-1,0) -- (1,0);
        \draw[thick] (-0.5,1) -- (0.5,1);
        \draw[thick] (-1,2) -- (1,2);
        \draw[thick] (0,0) -- (0,2);
      \end{scope}
      \node[anchor=north] at (5,-3) {$\bm{D_i}$};
      \begin{scope}[shift={(5,0)}]
        \draw[thick] (-1,0) -- (1,0);
        \draw[thick] (-0.5,0.8) -- (0.5,0.8);
        \draw[thick] (-0.5,1.6) -- (0.5,1.6);
        \draw[thick] (-1,2.4) -- (1,2.4);
        \draw[thick] (0,0) -- (0,2.4);
      \end{scope}
      \begin{scope}[shift={(5,2.4)}]
        \draw[thick] (-1,0) -- (1,0);
        \draw[thick] (-0.5,0.8) -- (0.5,0.8);
        \draw[thick] (-0.5,1.6) -- (0.5,1.6);
        \draw[thick] (-1,2.4) -- (1,2.4);
        \draw[thick] (0,0) -- (0,2.4);
      \end{scope}
      \begin{scope}[shift={(5,4.8)}]
        \draw[thick] (-1,0) -- (1,0);
        \draw[thick] (-0.5,0.8) -- (0.5,0.8);
        \draw[thick] (-0.5,1.6) -- (0.5,1.6);
        \draw[thick] (-1,2.4) -- (1,2.4);
        \draw[thick] (0,0) -- (0,2.4);
      \end{scope}
      \draw[thick,blue,->,>=stealth] (3.9,2.4) to [out=180,in=75] (2.7,3.6) to [out=255,in=0] (1.1,4.8);
      \draw[thick,black!30!green,->,>=stealth] (1.1,2.0) to [out=0,in=105] (2.3,3.4) to [out=285,in=180] (3.9,4.8);
    \end{tikzpicture}
    \caption{overlap}
    \label{fig:sched:overlap}
  \end{subfigure}%
  \begin{subfigure}{0.25\linewidth}
    \centering
    \begin{tikzpicture}[x=0.5em,y=-0.5em]
      \draw[white] (2.5,0) -- (2.5,8.6);
      \node[anchor=north] at (0,-3) {$\bm{G}$};
      \begin{scope}[shift={(0,0)}]
        \draw[thick] (-1,0) -- (1,0);
        \draw[thick] (-0.5,1) -- (0.5,1);
        \draw[thick] (-1,2) -- (1,2);
        \draw[thick] (0,0) -- (0,2);
      \end{scope}
      \begin{scope}[shift={(0,2)}]
        \draw[thick] (-1,0) -- (1,0);
        \draw[thick] (-0.5,1) -- (0.5,1);
        \draw[thick] (-1,2) -- (1,2);
        \draw[thick] (0,0) -- (0,2);
      \end{scope}
      \begin{scope}[shift={(0,4)}]
        \draw[thick] (-1,0) -- (1,0);
        \draw[thick] (-0.5,1) -- (0.5,1);
        \draw[thick] (-1,2) -- (1,2);
        \draw[thick] (0,0) -- (0,2);
      \end{scope}
      \begin{scope}[shift={(0,6)}]
        \draw[thick] (-1,0) -- (1,0);
        \draw[thick] (-0.5,1) -- (0.5,1);
        \draw[thick] (-1,2) -- (1,2);
        \draw[thick] (0,0) -- (0,2);
      \end{scope}
      \node[anchor=north] at (5,-3) {$\bm{D_i}$};
      \begin{scope}[shift={(5,0)}]
        \draw[thick] (-1,0) -- (1,0);
        \draw[thick] (-0.5,0.8) -- (0.5,0.8);
        \draw[thick] (-0.5,1.6) -- (0.5,1.6);
        \draw[thick] (-1,2.4) -- (1,2.4);
        \draw[thick] (0,0) -- (0,2.4);
      \end{scope}
      \begin{scope}[shift={(5,2.4)}]
        \draw[thick] (-1,0) -- (1,0);
        \draw[thick] (-0.5,0.8) -- (0.5,0.8);
        \draw[thick] (-0.5,1.6) -- (0.5,1.6);
        \draw[thick] (-1,2.4) -- (1,2.4);
        \draw[thick] (0,0) -- (0,2.4);
      \end{scope}
      \begin{scope}[shift={(5,4.8)}]
        \draw[thick] (-1,0) -- (1,0);
        \draw[thick] (-0.5,0.8) -- (0.5,0.8);
        \draw[thick] (-0.5,1.6) -- (0.5,1.6);
        \draw[thick] (-1,2.4) -- (1,2.4);
        \draw[thick] (0,0) -- (0,2.4);
      \end{scope}
      \draw[thick,blue,->,>=stealth] (3.9,2.4) to [out=180,in=45] (2.7,3.2) to [out=225,in=0] (1.1,4.0);
      \draw[thick,blue,->,>=stealth] (3.9,4.8) to [out=180,in=45] (2.7,5.4) to [out=225,in=0] (1.1,6.0);
      \draw[thick,black!30!green,->,>=stealth] (1.1,2.0) to [out=0,in=135] (2.3,2.6) to [out=315,in=180] (3.9,3.2);
      \draw[thick,black!30!green,->,>=stealth] (1.1,4.0) to [out=0,in=135] (2.3,4.8) to [out=315,in=180] (3.9,5.6);
    \end{tikzpicture}
    \caption{freerun}
    \label{fig:sched:freerun}
  \end{subfigure}
  \caption{Visualization of different training exchange schedules.  Generators shown performing two steps 
  (i.e. model updates) per exchange, while discriminators perform three.  Blue arrows represent discriminator model updates sent to generators, while green arrows show generator model updates sent to discriminators.}
  \label{fig:schedules}
\end{figure}

The {\em sequential} schedule (Figure~\ref{fig:sched:seq}) is equivalent to the dataflow of a purely
data-parallel approach.  The discriminator is trained for a specified number of steps, the updated discriminator
is used to train the generator, and so on.  As the figure suggests (and the results in
Section~\ref{sec:scaling-results} confirm), this schedule becomes inefficient once the GPUs are separated
into generator vs. discriminator workers, as one group or the other will always sit idle.  The {\em parallel} schedule (Figure~\ref{fig:sched:par})
addresses this by letting all workers train for a specified number of steps before exchanging models between
them.  However, it still exposes the latency of the model exchanges.  If slightly stale versions of the
adversarial models can be used without harming convergence (see Section~\ref{sec:schedconv}), the
{\em overlap} schedule (Figure~\ref{fig:sched:overlap}) overlaps the model exchange with training, leaving
any imbalance of work between the generator and discriminator workers as the remaining structural inefficiency.
The most aggressive {\em freerun} schedule (Figure~\ref{fig:sched:freerun}) attempts to eliminate this as well,
allowing each model to be trained continuously.  Model updates are sent periodically and applied by the
receiver on the next training step boundary.


\section{Performance Measurement}
\label{sec:measurement}

Although our networks are of feed forward type and are mainly comprised of fully connected layers which translate into matrix-matrix multiplication, the use of TensorFlow's automatic numeric differentiation to induce a network
for $f(x)$ makes it difficult to compute the number of floating point operations required for a training step
directly from the layer parameters.  Instead, we traverse the TensorFlow operation graph after all transformations
have been performed (i.e. both the explicit differentiation for the PDE and the automatic differentiation
necessary to support backpropagation) and estimate the operations required for each node in the graph.  Since the
operation graphs differ from GPU to GPU (depending on the size of its assigned subdomain and whether it is training
the discriminator or the generator), each GPU's workload must be estimated independently.
To validate these estimates, we used the API logging capability of the cuBLAS and cuDNN libraries to
generate a log file listing every call made into either library, including all parameter values.  These logs
were then parsed to compute independent estimates of the FLOPs required for a training step of each
generator and discriminator network, which were found to agree with the in-application estimate.
In order to extract FLOP rates, we combine this information with time stamps printed by each group of networks for each iteration. Since for most training schedules our code is predominantly asynchronous, those time steps generally do not align across groups. We therefore interpolate the time series for each group individually and align them by resampling in 1 ms intervals, which corresponds to about 1/3 of the time for a discriminator iteration. We then compute the aggregate FLOPs by summing across all groups for each individual time interval and divide by the interval length to get the FLOP rate. We finally compute the median and the maximum across all time intervals and define this as our sustained and peak FLOP rates, respectively. For visualizing performance variations, we further compute the 68\% confidence levels (i.e. ${\sim}1$ standard deviation) for each concurrency and show those as asymmetric error bars in our scaling plots presented in Section \ref{sec:scaling-results}.

\subsection{HPC System and Environment}

\subsubsection{Hardware configuration}
\label{sec:summit} 
All of our experiments were performed on the Summit, a leadership class supercomputer installed at the Oak Ridge National Laboratory (ORNL).  Summit is currently the
world's fastest system according to the biannual TOP500 ranking~\cite{top500-summit-site}.  
The system is comprised of 4608 nodes, each equipped with two IBM Power 9 CPUs and 6 NVIDIA Volta GPUs (V100) with 16 GB HBM2 memory.
Each Power 9 CPU is connected to 3 Volta GPUs using NVIDIA high-speed interconnect NVLink, capable of  300 GB/s bi-directional bandwidth.
For inter-node communication, each node uses a pair of dual-rail EDR Infiniband cards capable of injecting 23 GB/s into a non-blocking fat-tree topology.

The Volta GPU architecture incorporates Tensor Cores that accelerate mixed-precision operations common to deep
learning (and many other) workloads.
Each of the 640 Tensor Cores can perform 64 floating-point Fused-Multiply-Add (FMA) operations per cycle, for
a peak throughput of 125 TF/s per GPU (750 TF/s per Summit node).
Input operands are provided in half precision (FP16), while outputs may be either half or single precision.

\subsubsection{Software environment}
Our experiments were performed using TensorFlow v1.12, compiled with CUDA 9.2.148, cuDNN 7.5.0, and cuBLAS 9.2.174.  Internode model exchanges (see Figure~\ref{fig:schedules}) used IBM Spectrum MPI v10.2.0.10, while gradient reductions during training were performed with Horovod v0.15.2, compiled with NCCL v2.4.6.

\section{Results}
\label{sec:results}

\subsection{Physics-informed GAN Results}
\label{sec:science_results}


\subsubsection{Correctness of solution for 1 domain}

\begin{figure}[h]
  \includegraphics[width=\linewidth]{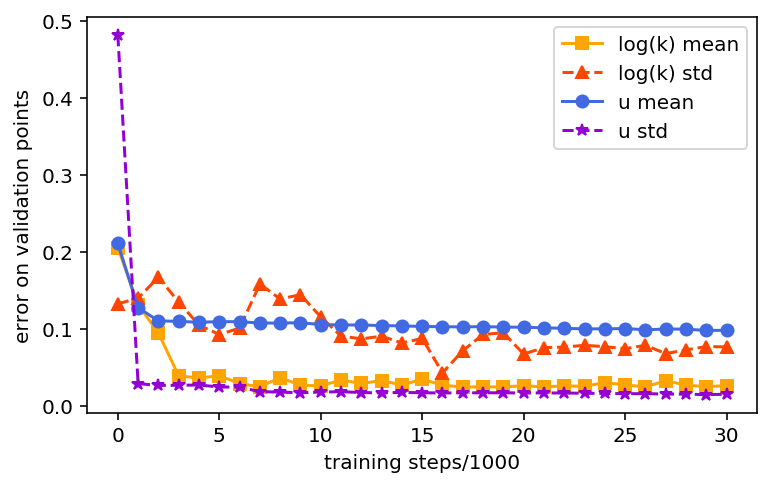}
  \caption{Error of mean and standard deviation for $\log(k)$, $u$ on validation points.}
  \label{fig:cartoon1}
\end{figure}

\begin{figure}[h]
  \includegraphics[width=\linewidth]{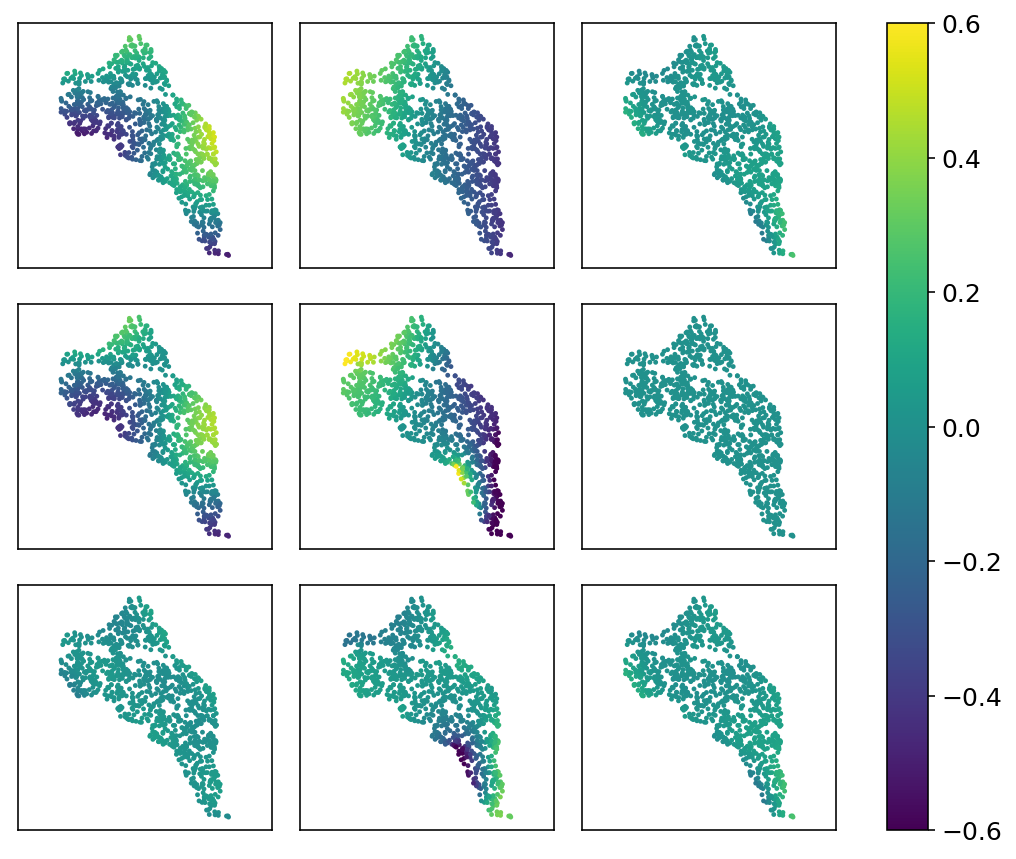}
  \caption{Validation of GAN predictions. The columns represent the mean of $\log k$ (left), $u$ (middle) and $f$ (right) fields. The top row shows the GAN prediction, the middle row the ground truth and the bottom row shows the difference between these two.}
  \label{fig:cartoon2}
\end{figure}

\begin{figure}[h]
  \includegraphics[width=\linewidth]{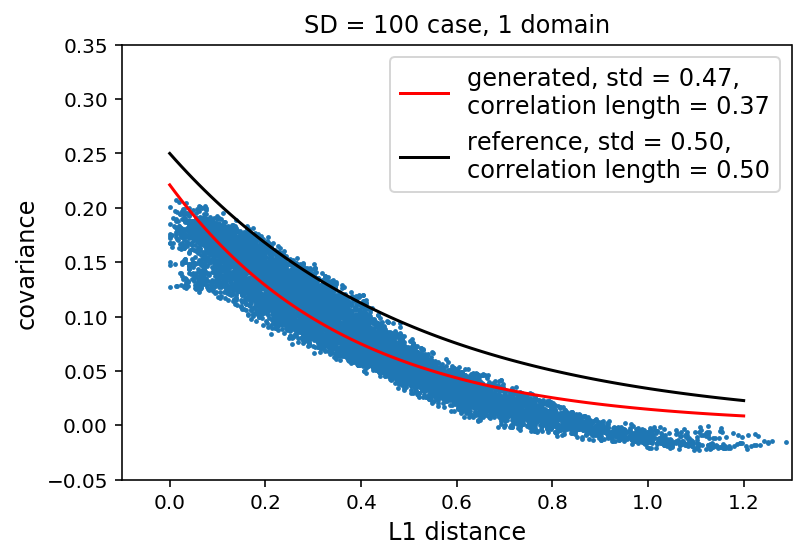}
  \caption{Comparison of covariance and $L_1$ distance of $\log(k)$ learned by PI-GAN. Blue points indicate covariance between PI-GAN generated data.}
  \label{fig:cartoon3}
\end{figure}

We train the PI-GANs with Dataset \#3 level 1 to validate the correctness of solution for 1 domain that covers the entire Hanford Site. Figures \ref{fig:cartoon1} \ref{fig:cartoon2}, and \ref{fig:cartoon3} show that the GAN makes realistic predictions of $\log(k)$ field and $u$ field.
The errors for $u$ are larger at the boundaries as we have not incorporated the boundary conditions directly into our current PI-GAN implementation.

\subsubsection{Correctness of solution for multiple sub-domains and high stochastic dimensionality}
\begin{figure}[h]
  \includegraphics[width=\linewidth]{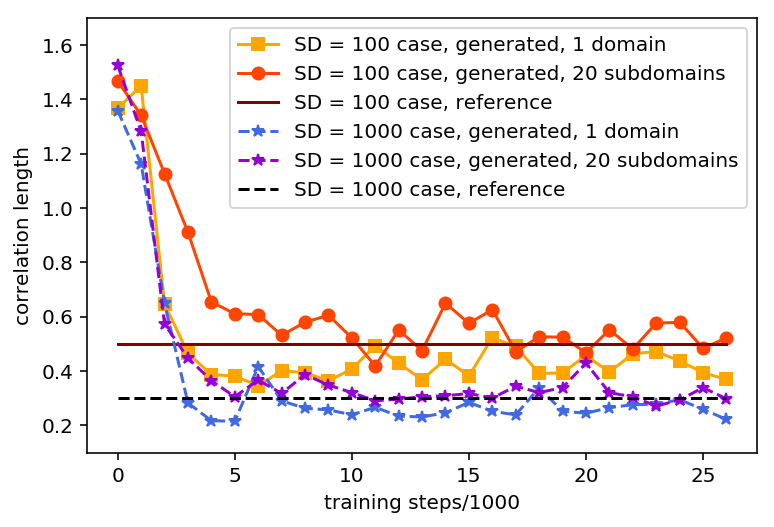}
  \caption{Correlation length of $\log(k)$ learned by PI-GAN for SD = 100 and SD = 1000, 1 domain and 20 subdomains.}
  \label{fig:domSD1}
\end{figure}

\begin{figure}[h]
  \includegraphics[width=\linewidth]{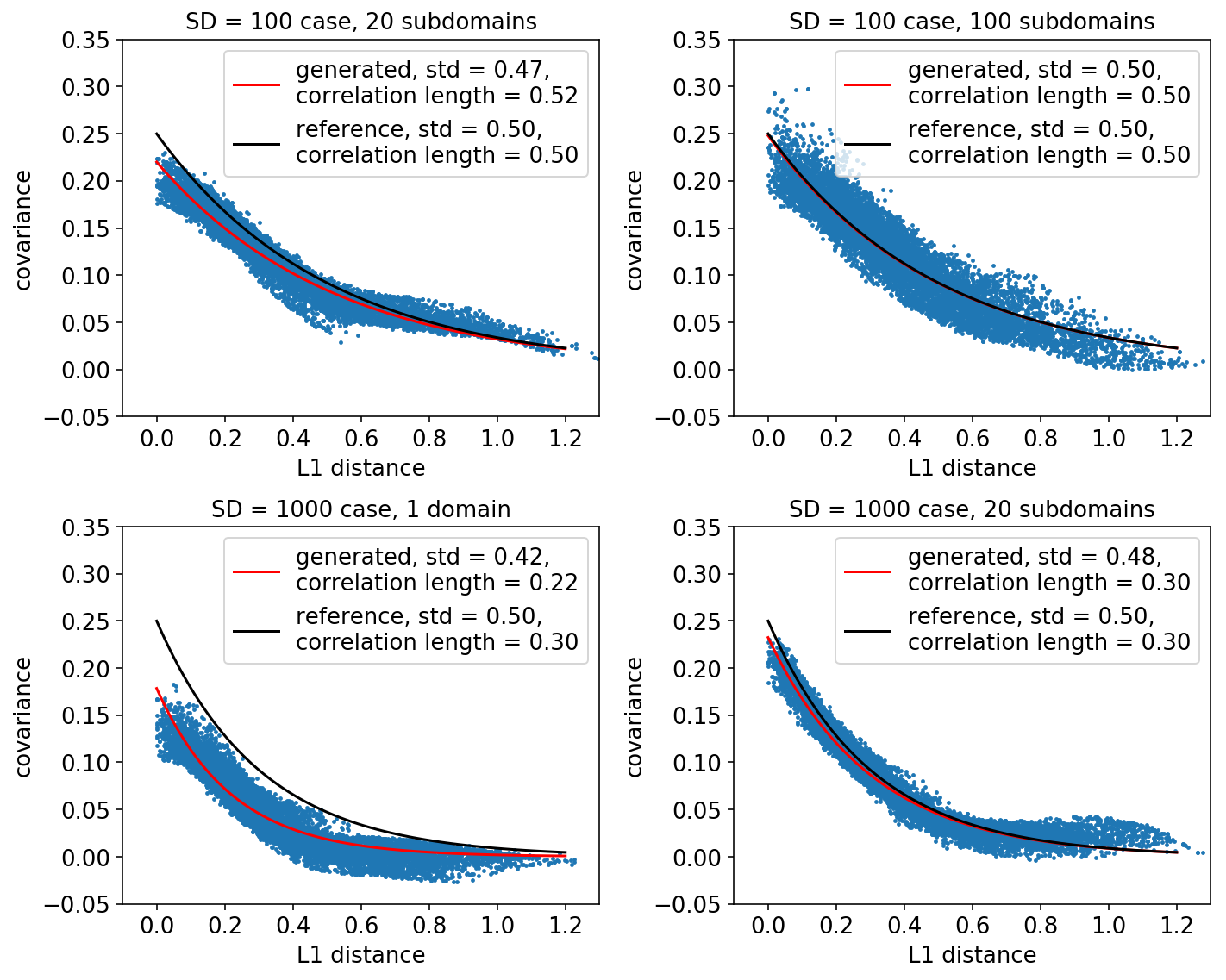}
  \caption{Comparison of covariance and $L_1$ distance of $\log(k)$ learned by PI-GAN as we vary number of subdomains and stochastic dimensionality. Blue points indicate covariance between PI-GAN generated data.}
  \label{fig:domSD2}
\end{figure}

Next we present results for PI-GANs for 1, 20 and 100 subdomains, as well as  SD = 100 and 1000. 
For the case of SD = 100, we train PI-GANs with Dataset \#3 level 1 (1 domain), Dataset \#3 level 1 and 2 (20 subdomains in all), Dataset \#2 (100 subdomains in all). For the case of SD = 1000, we train PI-GANs with Dataset \#4 level 1 (1 domain), and Dataset \#4 level 1 and 2 (20 subdomains in all).
Figure \ref{fig:domSD1} shows errors for the quantities of interest for 100 and 1,000 stochastic dimensions (SD) for 1 and 20 domains; the results for 100 domains are similar. Figure
\ref{fig:domSD2} shows the pairwise covariance between 100 validation points scattered randomly across the Hanford Site. The red line, which is the best-fit to the PI-GANs generated $\log(k)$ field, is in agreement with the reference (black line) values computed from the exact covariance kernel. We can see clearly the inadequacy of the model using a single domain that with only 100 sensors (lower left in Figure~\ref{fig:domSD2}) compared to the multidomain cases. The comprehensive diagnostic results shown here validate our choice of PI-GAN architecture. 

\subsubsection{Effect of training schedules and fp16 vs. fp32}
\label{sec:schedconv}
\begin{figure}[h]
  \includegraphics[width=\linewidth]{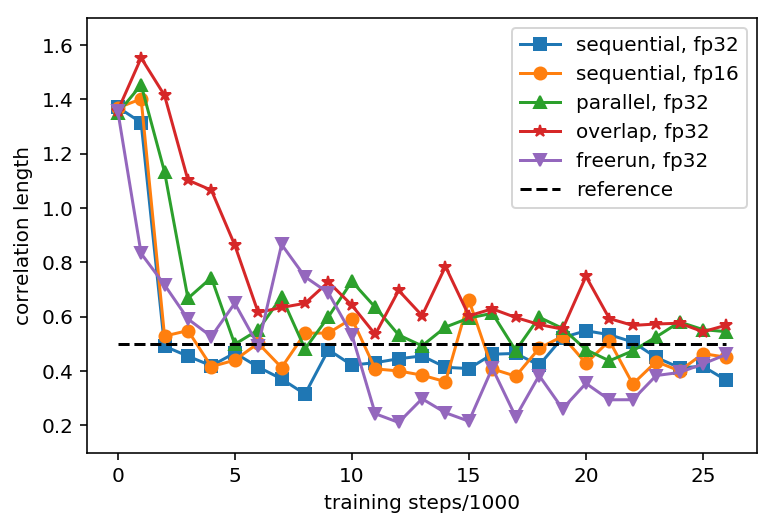}
  \caption{Correlation length of $\log(k)$ learned by PI-GAN with different training schedules and fp16 vs. fp32.}
  \label{fig:sched1}
\end{figure}

\begin{figure}[h]
  \includegraphics[width=\linewidth]{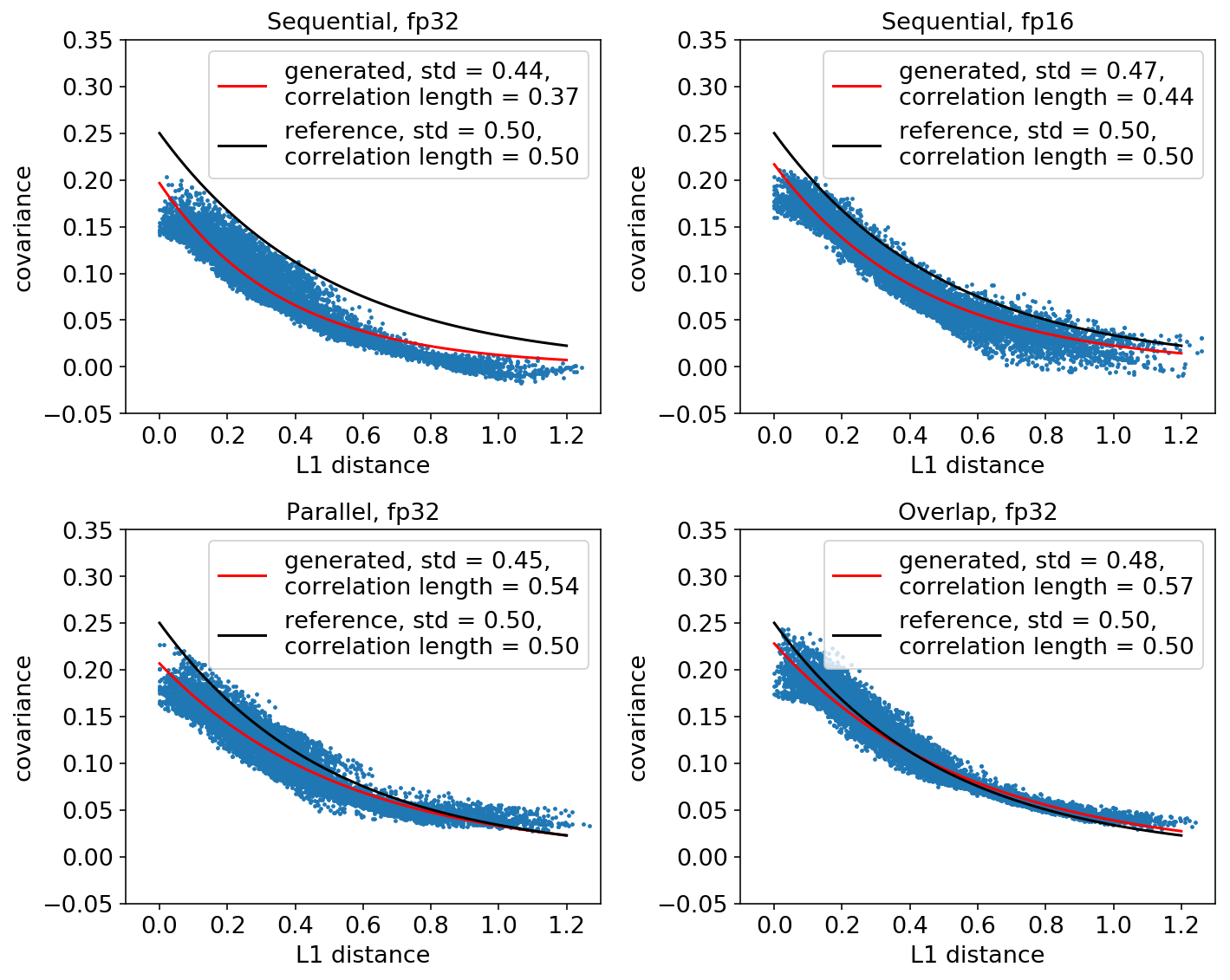}
  \caption{Comparison of covariance and $L_1$ distance of $\log(k)$ learned by PI-GAN for various training schedules and fp16 vs. fp32. Blue points indicate covariance between PI-GAN generated data.}
  \label{fig:sched2}
\end{figure}

In order to investigate the effect of training schemes presented in Figure \ref{fig:schedules} as well as the reduced arithmetic precision employed, we make explicit comparisons of the results for the case of Dataset \#1
in Figures~\ref{fig:sched1} and \ref{fig:sched2}. 
The mixed-precision training is at least as good as the single-precision results in terms of both convergence rate
and the accuracy of the resulting model.  Comparisons between single- and mixed-precision for other datasets,
schedules, and scales yielded no meaningful differences, and the corresponding plots have been omitted as a result.

The behavior of the different model exchange schedules is a little more
varied.  The parallel schedule's convergence and quality is indistinguishable from that of the sequential schedule,
allowing us to adopt the parallel schedule as the baseline for large scaling runs.  The more aggressive
overlap and freerun schedules show
excellent model accuracy in Figure~\ref{fig:sched2}, but convergence appears to be slower.  As with any
hyperparameter, the choice of schedule needs to be re-examined at each scale to determine the best trade-off between
step time and the total number of steps needed for convergence.


\subsection{Single Node Performance}
\label{sec:perfsingle}
To examine the per-GPU performance before communication and scaling issues are taken into consideration, we
performed tests using 2 GPUs on a single Summit node for just the top-level subdomain in the largest dataset.
One GPU was therefore the (sole) generator worker and the other the discriminator worker.  The parallel schedule
was used with a large number of steps per exchange to minimize the timing impact of the model exchange.  The results shown in Table~\ref{tab:singlegpu} show the batch size, number of operations per step,
and single-GPU throughput for each GPU separately.  The final column is a weighted average of the generator
and discriminator performance, estimating the expected average performance per GPU for larger jobs.  (This is
deviates from a simple average as the relative performance of generator and discriminator workers changes.)
Single-precision training throughput is 12.5 TF/s per GPU, with the discriminators operating over 45\% faster
than the generators, most likely due to the larger batch size.
For mixed-precision training, which is able to take advantage of the GPU's Tensor Cores, performance improves
by $3.7\times$ over the single-precision throughput.  The generator performance per GPU is {\generatorperf} TF/s while the discriminator
performance is {\discriminatorperf} TF/s, for an effective average performance of 49.0 TF/s per GPU.

In
addition to examining the network as designed with the {\em tanh} activation function, we also measured the
performance when using the simpler {\em relu} activation function to understand the potential performance
improvement available if the differentiability concerns described in Section~\ref{sec:innov:singlenode} can
be resolved.
With a potential benefit of over 16\% from switch to {\em relu}, further study on its use is warranted.
However, all of the other results shown use the {\em tanh} activation function.

\begin{table}[t]
\resizebox{\linewidth}{!}{
\begin{tabular}{|ll|ccc|ccc|c|}
\hline
& & \multicolumn{3}{c|}{Generator} & \multicolumn{3}{c|}{Discriminator} & Effective \\
Prec. & Act. & Batch & FLOP/ & Perf. & Batch & FLOP/ & Perf. & Performance \\
& Func & Size & step & & Size & step & & \\
& & & ($\times 10^{12}$) & \textbf{(TF/s)} & & ($\times 10^{12}$) & \textbf{(TF/s)} & \textbf{(TF/s)} \\ \hline
FP16 & tanh & 6 & 12.3 & 38.6 & 32 & 21.9 & 59.3 & 49.0 \\
& relu & & & 52.7 & & & 60.8 & 57.3 \\ \hline
FP32 & tanh & 3 & 6.15 & 10.1 & 16 & 10.9 & 14.7 & 12.5 \\
& relu & & & 11.7 & & & 14.4 & 13.2 \\ \hline
\end{tabular}
}
\caption{Single-GPU performance of PI-GAN network.  Results shown for both half precision (FP16) and single precision (FP32) training.  The effective performance is a weighted average that takes into account the ratio of generator workers to discriminator workers.  The {\em tanh} activation function was used for all other experiments, but the potential performance improvement of switching to {\em relu} is shown here.}\label{tab:singlegpu}
\end{table}

\subsection{Multi Node Scaling}
\label{sec:scaling-results}

\begin{figure}[h]
  \includegraphics[width=\linewidth]{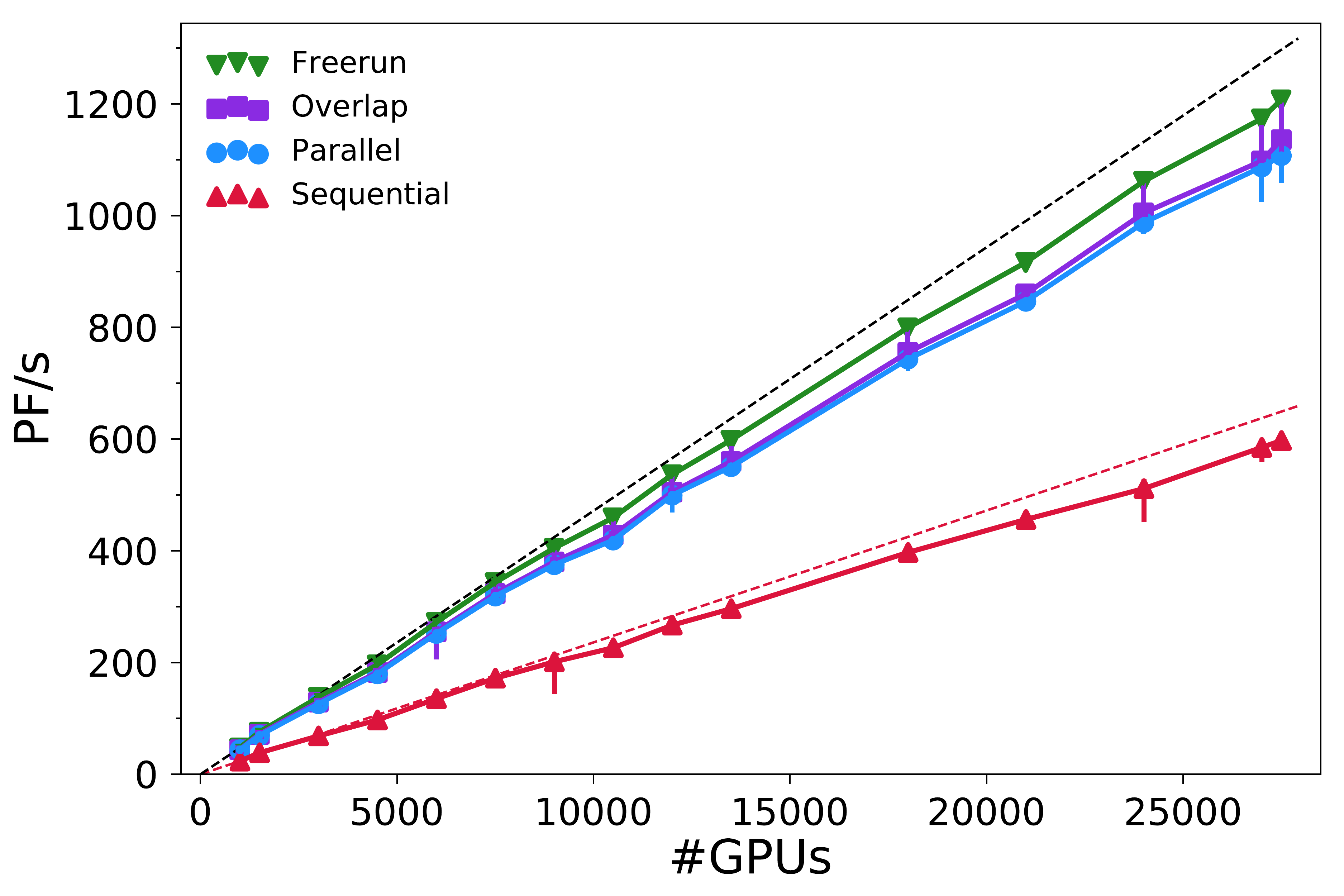}
  \caption{Performance scaling for sequential (red), parallel (blue), overlap (purple) and freerun (green) training schedules. The red dashed line represent ideal scaling for the sequential and the black dashed line for the parallel training variants. The ideal performance is extrapolated from the observed performance on the smallest possible scale which could fit the problem, i.e. 1000 GPUs.}
  \label{fig:scaling}
\end{figure}

In this section we explore the scalability of the largest of the three networks shown in Figure~\ref{fig:cartoon},
which is the largest network the Summit system can support and is intended to model the Hanford Site with a stochastic dimensionality of at least 10,000.  We note that a 
careful (but expensive!) analysis might be able to show a smaller network is sufficient for 10,000 dimensions.
However, such a result would increase the maximum dimensionality of configurations that can be modeled on
Summit --- the
scalability of the large network remains relevant.
Dataset \#6 (in Table~\ref{tab:datasets})
contains 500 subdomains, each of which has 2,200 sensors (100 k, 100 u, and 2000 f) sensors and 1,000,000 realizations for a total of 17.6 TB of training data.  As each subdomain utilizes (at least)
one GPU for discriminator training and one for generator training, the application must be be run on at least
1000 GPUs in total (${\sim}167$ Summit nodes) and we test its performance scaling behavior up to {\herogpucount}  GPUs (nearly the entire Summit system).

The neural network configuration used for this enormous problem is $21{\times}1024$, and we use local batch sizes of
32 for discriminator training and 6 for generator training, both limited by the memory capacity of the Volta GPUs.
Figure~\ref{fig:scaling} shows the performance scaling behavior for each of the four schedules described in
Section~\ref{sec:schedules}.  The black dotted line represents ideal scaling for the freerun, overlap, and
parallel schedules, while the red dotted line represents the (significantly lower) ideal scaling for the 
sequential schedule.  At {\herogpucount} GPUs ({\heronodecount} Summit nodes), the sequential schedule achieves a peak performance of {\heropeakperfseq} PF/s and a sustained performance of {\herosustainedperfseq} PF/s, for a parallel efficiency of {\weakscalingeffseq}\%.  The parallel schedule, which is able to use all GPUs concurrently, reaches a peak and sustained performances of {\heropeakperfpar} PF/s and {\herosustainedperfpar} PF/s respectively.  This represents a parallel efficiency of {\weakscalingeffpar}\% and a
$1.87\times$ improvement over the sequential schedule.  The overlap schedule improves to
{\heropeakperfover} PF/s (peak) and {\herosustainedperfover} PF/s (sustained), for a parallel efficiency of {\weakscalingeffover}\%.  The freerun schedule further
improves the sustained performance on {\herogpucount} GPUs to {\herosustainedperf} PF/s with a peak performance
of {\heropeakperf} PF/s and a parallel efficiency of {\weakscalingeff}\%.

The scaling plots in Figure~\ref{fig:scaling} do show a few minor anomalies in the overall scaling trend.
Our work distribution strategy has been designed to balance generator and discriminator training in the adversarial model,
but quantization effects can skew this balance for some points on the scaling curve.  For example, when run on a
total of 10,500 GPUs, each subdomain is assigned 10 GPUs for generator training and 11 for discriminator training,
causing an efficiency loss of a few percent.

We note that Dataset \#6 has a correlation length that is short relative to the Hanford Site length-scale, which induces significant small-scale features in the realizations of the $\log(k)$ field.
The performance scaling results for Dataset \#6 indicate that the current PI-GAN implementation is fully capable of processing the amount of data required to solve the mixed SPDE problem in the regime of high dimensionality.



\subsection{Discussion}
Solving SPDEs in high dimensions and tackling the curse of dimensionality is a holy grail in computational science, and the proposed PI-GAN approach has a potential to contribute greatly towards this goal.
Such foundational contribution will be the keystone in building a "digital twin" for the critical region of the Hanford Site and will become a paradigm for similar digital twins of complex multiscale problems that require continuous monitoring via data and simulations.
However, here we only considered a subset of the total multiphysics problem that involves a coupled system of SPDEs.
Future work should address the development of more sophisticated DNNs and GANs that best represent the multiscale and multiphysics nature of these physical twins with proper incorporation of boundary and interface conditions, external interactions (e.g., rain), biochemistry, etc.

Future work should also address current limitations of the GANs architecture proposed in this paper.
Although we have introduced domain decomposition for discriminators, we only have one generator for the whole domain.
In the future we will explore GANs structure with separate generator assigned to each subdomain, where the generators will be synchronized across subdomain boundaries via continuity constraints stemming from the physics.


\section{Conclusions}
\label{sec:conclusions}
In this work, we have taken an important step towards bridging the distinct worlds of theory-driven modeling, and data-driven analysis. Motivated by the important problem of uncertainty quantification for characterizing subsurface flow at the Hanford Site, we have proposed a state-of-the-art GAN model that is informed by a stochastic PDE constraint. Our model is able to process data across multiple subdomains, and is effectively mapped to execute on multiple GPUs. We implement our application in the high-level TensorFlow framework, and obtain high performance levels (\generatorperf-\discriminatorperf   TF/s) on Volta GPUs. Building upon a number of communication and load balancing strategies, we were successful in scaling the model out to \herogpucount GPUs, obtaining a peak and sustained performance of \heropeakperf  PF/s and \herosustainedperf  PF/s respectively. To the best of our knowledge, this is the first exascale-ready implementation of a GAN architecture and the first successful attempt at solving mixed stochastic PDEs with 1,000 (or higher) dimensionality. 


\section*{Acknowledgments}
This research used resources of the National Energy Research Scientific Computing Center (NERSC), a DOE Office of Science User Facility supported by the Office of Science of the U.S. Department of Energy under Contract No. DE-AC02-05CH11231. This research used the Summit system at the Oak Ridge Leadership Computing Facility at the Oak Ridge National Laboratory, which is supported by the Office of Science of the U.S. Department of Energy under Contract No. DE-AC05-00OR22725. At the Pacific   Northwest   National Laboratory (PNNL), this  work  was  partially  supported  by  the Laboratory Directed Research and Development Program (LDRD) and the U.S. Department of Energy (DOE) Office of Science, Office of Advanced Scientific Computing Research (ASCR) project. PNNL is operated by Battelle for the DOE under Contract DE-AC05-76RL01830. At Brown University, this work was supported by the Department of Energy (DOE) grants DE-SC0019434 and DE-SC0019453.


\bibliographystyle{IEEEtran}
\bibliography{IEEEabrv,main}

\begin{thebibliography}{10}
\providecommand{\url}[1]{#1}
\csname url@samestyle\endcsname
\providecommand{\newblock}{\relax}
\providecommand{\bibinfo}[2]{#2}
\providecommand{\BIBentrySTDinterwordspacing}{\spaceskip=0pt\relax}
\providecommand{\BIBentryALTinterwordstretchfactor}{4}
\providecommand{\BIBentryALTinterwordspacing}{\spaceskip=\fontdimen2\font plus
\BIBentryALTinterwordstretchfactor\fontdimen3\font minus
  \fontdimen4\font\relax}
\providecommand{\BIBforeignlanguage}[2]{{%
\expandafter\ifx\csname l@#1\endcsname\relax
\typeout{** WARNING: IEEEtran.bst: No hyphenation pattern has been}%
\typeout{** loaded for the language `#1'. Using the pattern for}%
\typeout{** the default language instead.}%
\else
\language=\csname l@#1\endcsname
\fi
#2}}
\providecommand{\BIBdecl}{\relax}
\BIBdecl

\bibitem{thorne2006groundwater}
P.~D. Thorne, M.~P. Bergeron, M.~D. Williams, and V.~L. Freedman, ``Groundwater
  data package for hanford assessments,'' Pacific Northwest National
  Lab.(PNNL), Richland, WA (United States), Tech. Rep., 2006.

\bibitem{MaziarParisGK17-1}
M.~{Raissi}, P.~{Perdikaris}, and G.~E. {Karniadakis}, ``{Physics Informed Deep
  Learning (Part I): Data-driven Solutions of Nonlinear Partial Differential
  Equations},'' \emph{arXiv e-prints}, p. arXiv:1711.10561, Nov 2017.

\bibitem{MaziarParisGK17-2}
------, ``{Physics Informed Deep Learning (Part II): Data-driven Discovery of
  Nonlinear Partial Differential Equations},'' \emph{arXiv e-prints}, p.
  arXiv:1711.10566, Nov 2017.

\bibitem{YangLiu-GAN}
L.~{Yang}, D.~{Zhang}, and G.~E. {Karniadakis}, ``{Physics-Informed Generative
  Adversarial Networks for Stochastic Differential Equations},'' \emph{arXiv
  e-prints}, p. arXiv:1811.02033, Nov 2018.

\bibitem{wasserstein1997monte}
G.~S. Fishman, ``Monte carlo: Concepts, algorithms, and applications,''
  \emph{Technometrics}, vol.~39, no.~3, pp. 338--338, 1997.

\bibitem{wiener1938homogeneous}
\BIBentryALTinterwordspacing
N.~Wiener, ``The homogeneous chaos,'' \emph{American Journal of Mathematics},
  vol.~60, no.~4, pp. 897--936, 1938. [Online]. Available:
  \url{http://www.jstor.org/stable/2371268}
\BIBentrySTDinterwordspacing

\bibitem{XiuKarniadakis-2003}
\BIBentryALTinterwordspacing
D.~Xiu and G.~E. Karniadakis, ``Modeling uncertainty in flow simulations via
  generalized polynomial chaos,'' \emph{Journal of Computational Physics}, vol.
  187, no.~1, pp. 137 -- 167, 2003. [Online]. Available:
  \url{http://www.sciencedirect.com/science/article/pii/S0021999103000925}
\BIBentrySTDinterwordspacing

\bibitem{elman2011assessment}
H.~C. Elman, C.~W. Miller, E.~T. Phipps, and R.~S. Tuminaro, ``Assessment of
  collocation and galerkin approaches to linear diffusion equations with random
  data,'' \emph{International Journal for Uncertainty Quantification}, vol.~1,
  no.~1, pp. 19--33, 2011.

\bibitem{JasmineFoo}
\BIBentryALTinterwordspacing
J.~Foo and G.~E. Karniadakis, ``Multi-element probabilistic collocation method
  in high dimensions,'' \emph{Journal of Computational Physics}, vol. 229,
  no.~5, pp. 1536 -- 1557, 2010. [Online]. Available:
  \url{http://www.sciencedirect.com/science/article/pii/S0021999109006044}
\BIBentrySTDinterwordspacing

\bibitem{Handy-SISC}
\BIBentryALTinterwordspacing
Z.~Zhang, M.~Choi, and G.~Karniadakis, ``Error estimates for the anova method
  with polynomial chaos interpolation: Tensor product functions,'' \emph{SIAM
  Journal on Scientific Computing}, vol.~34, no.~2, pp. A1165--A1186, 2012.
  [Online]. Available: \url{https://doi.org/10.1137/100788859}
\BIBentrySTDinterwordspacing

\bibitem{barajassolano-2016-stochastic}
D.~A. Barajas-Solano and D.~M. Tartakovsky, ``Stochastic collocation methods
  for nonlinear parabolic equations with random coefficients,'' \emph{SIAM/ASA
  J. Uncert. Quantif.}, vol.~4, pp. 475--494, 2016.

\bibitem{evensen2003ensemble}
G.~Evensen, ``The ensemble kalman filter: Theoretical formulation and practical
  implementation,'' \emph{Ocean dynamics}, vol.~53, no.~4, pp. 343--367, 2003.

\bibitem{Maziar-GPR}
\BIBentryALTinterwordspacing
M.~Raissi and G.~E. Karniadakis, ``Hidden physics models: Machine learning of
  nonlinear partial differential equations,'' \emph{Journal of Computational
  Physics}, vol. 357, pp. 125 -- 141, 2018. [Online]. Available:
  \url{http://www.sciencedirect.com/science/article/pii/S0021999117309014}
\BIBentrySTDinterwordspacing

\bibitem{Maziar-PINNs}
\BIBentryALTinterwordspacing
M.~Raissi, P.~Perdikaris, and G.~Karniadakis, ``Physics-informed neural
  networks: A deep learning framework for solving forward and inverse problems
  involving nonlinear partial differential equations,'' \emph{Journal of
  Computational Physics}, vol. 378, pp. 686 -- 707, 2019. [Online]. Available:
  \url{http://www.sciencedirect.com/science/article/pii/S0021999118307125}
\BIBentrySTDinterwordspacing

\bibitem{tartakovsky2018learning}
A.~M. {Tartakovsky}, C.~{Ortiz Marrero}, P.~{Perdikaris}, G.~D. {Tartakovsky},
  and D.~{Barajas-Solano}, ``{Learning Parameters and Constitutive
  Relationships with Physics Informed Deep Neural Networks},'' \emph{arXiv
  e-prints}, p. arXiv:1808.03398, Aug 2018.

\bibitem{Weinan-arxiv}
W.~{E}, J.~{Han}, and A.~{Jentzen}, ``{Deep learning-based numerical methods
  for high-dimensional parabolic partial differential equations and backward
  stochastic differential equations},'' \emph{arXiv preprint}, p.
  arXiv:1706.04702, jun 2017.

\bibitem{Maziar-arxiv}
M.~{Raissi}, ``{Forward-backward stochastic neural networks: Deep learning of
  high-dimensional partial differential equations},'' \emph{arXiv preprint}, p.
  arXiv:1804.07010, apr 2018.

\bibitem{Paris-encoder}
Y.~{Yang} and P.~{Perdikaris}, ``{Adversarial Uncertainty Quantification in
  Physics-Informed Neural Networks},'' \emph{arXiv e-prints}, p.
  arXiv:1811.04026, Nov 2018.

\bibitem{zhang2018quantifying}
D.~{Zhang}, L.~{Lu}, L.~{Guo}, and G.~E. {Karniadakis}, ``{Quantifying total
  uncertainty in physics-informed neural networks for solving forward and
  inverse stochastic problems},'' \emph{arXiv e-prints}, p. arXiv:1809.08327,
  Sep 2018.

\bibitem{goodfellow2014generative}
I.~Goodfellow, J.~Pouget-Abadie, M.~Mirza, B.~Xu, D.~Warde-Farley, S.~Ozair,
  A.~Courville, and Y.~Bengio, ``Generative adversarial nets,'' in
  \emph{Advances in neural information processing systems (2014)}, 2014, pp.
  2672--2680.

\bibitem{dcgan-paper-2015}
A.~{Radford}, L.~{Metz}, and S.~{Chintala}, ``{Unsupervised Representation
  Learning with Deep Convolutional Generative Adversarial Networks},''
  \emph{arXiv e-prints}, p. arXiv:1511.06434, Nov 2015.

\bibitem{stylegan}
T.~{Karras}, S.~{Laine}, and T.~{Aila}, ``{A Style-Based Generator Architecture
  for Generative Adversarial Networks},'' \emph{arXiv e-prints}, p.
  arXiv:1812.04948, Dec 2018.

\bibitem{arjovsky2017wasserstein}
M.~Arjovsky, S.~Chintala, and L.~Bottou, ``Wasserstein {GAN},'' \emph{arXiv
  preprint}, p. arXiv:1701.07875, 2017.

\bibitem{gulrajani2017improved}
I.~Gulrajani, F.~Ahmed, M.~Arjovsky, V.~Dumoulin, and A.~C. Courville,
  ``Improved training of {Wasserstein GANs},'' in \emph{Advances in Neural
  Information Processing Systems (2017)}, 2017, pp. 5767--5777.

\bibitem{largegan}
\BIBentryALTinterwordspacing
A.~Brock, J.~Donahue, and K.~Simonyan, ``Large scale {GAN} training for high
  fidelity natural image synthesis,'' \emph{CoRR}, vol. abs/1809.11096, 2018.
  [Online]. Available: \url{http://arxiv.org/abs/1809.11096}
\BIBentrySTDinterwordspacing

\bibitem{cole2001uncertainty}
C.~R. Cole, M.~P. Bergeron, C.~J. Murray, P.~D. Thorne, S.~K. Wurstner, and
  P.~M. Rogers, ``Uncertainty analysis framework-hanford site-wide groundwater
  flow and transport model,'' Pacific Northwest National Lab., Richland, WA
  (US), Tech. Rep., 2001.

\bibitem{karypis1998metis}
\BIBentryALTinterwordspacing
G.~Karypis and V.~Kumar, ``A fast and high quality multilevel scheme for
  partitioning irregular graphs,'' \emph{SIAM Journal on Scientific Computing},
  vol.~20, no.~1, pp. 359--392, 1998. [Online]. Available:
  \url{https://doi.org/10.1137/S1064827595287997}
\BIBentrySTDinterwordspacing

\bibitem{tensorflow2015-whitepaper}
\BIBentryALTinterwordspacing
M.~Abadi, A.~Agarwal, P.~Barham, E.~Brevdo, Z.~Chen, C.~Citro, G.~S. Corrado,
  A.~Davis, J.~Dean, M.~Devin, S.~Ghemawat, I.~Goodfellow, A.~Harp, G.~Irving,
  M.~Isard, Y.~Jia, R.~Jozefowicz, L.~Kaiser, M.~Kudlur, J.~Levenberg,
  D.~Man\'{e}, R.~Monga, S.~Moore, D.~Murray, C.~Olah, M.~Schuster, J.~Shlens,
  B.~Steiner, I.~Sutskever, K.~Talwar, P.~Tucker, V.~Vanhoucke, V.~Vasudevan,
  F.~Vi\'{e}gas, O.~Vinyals, P.~Warden, M.~Wattenberg, M.~Wicke, Y.~Yu, and
  X.~Zheng, ``{TensorFlow}: Large-scale machine learning on heterogeneous
  systems,'' 2015, software available from tensorflow.org. [Online]. Available:
  \url{https://www.tensorflow.org/}
\BIBentrySTDinterwordspacing

\bibitem{tensorflow-site}
\BIBentryALTinterwordspacing
Google. (2019) {TensorFlow website}. [Online]. Available:
  \url{https://tensorflow.org}
\BIBentrySTDinterwordspacing

\bibitem{resnet}
K.~He, X.~Zhang, S.~Ren, and J.~Sun, ``Deep residual learning for image
  recognition,'' in \emph{{CVPR}}.\hskip 1em plus 0.5em minus 0.4em\relax
  {IEEE} Computer Society, 2016, pp. 770--778.

\bibitem{nvidia:mixedprectraining}
\BIBentryALTinterwordspacing
Nvidia, ``Training with mixed precision,'' April 2019. [Online]. Available:
  \url{https://docs.nvidia.com/deeplearning/sdk/pdf/Training-Mixed-Precision-User-Guide.pdf}
\BIBentrySTDinterwordspacing

\bibitem{relu}
\BIBentryALTinterwordspacing
A.~F. Agarap, ``Deep learning using rectified linear units (relu),''
  \emph{CoRR}, vol. abs/1803.08375, 2018. [Online]. Available:
  \url{http://arxiv.org/abs/1803.08375}
\BIBentrySTDinterwordspacing

\bibitem{sergeev2018horovod}
A.~Sergeev and M.~D. Balso, ``Horovod: fast and easy distributed deep learning
  in {TensorFlow},'' \emph{arXiv preprint arXiv:1802.05799}, 2018.

\bibitem{Tencent}
X.~{Jia}, S.~{Song}, W.~{He}, Y.~{Wang}, H.~{Rong}, F.~{Zhou}, L.~{Xie},
  Z.~{Guo}, Y.~{Yang}, L.~{Yu}, T.~{Chen}, G.~{Hu}, S.~{Shi}, and X.~{Chu},
  ``{Highly Scalable Deep Learning Training System with Mixed-Precision:
  Training ImageNet in Four Minutes},'' \emph{ArXiv e-prints}, Jul. 2018.

\bibitem{top500-summit-site}
\BIBentryALTinterwordspacing
O.~R.~N. Laboratory. (2019) {Summit - IBM Power System AC922, IBM POWER9 22C
  3.07GHz, NVIDIA Volta GV100, Dual-rail Mellanox EDR Infiniband | TOP500
  Supercomputer Sites}. [Online]. Available:
  \url{https://www.top500.org/system/179397}
\BIBentrySTDinterwordspacing

\end{thebibliography}

\end{document}